\documentclass{acmlarge}
\usepackage[utf8]{inputenc}
\makeatletter
\def\runningfoot{\def\@runningfoot{}}
\def\@runningfoot{}
\def\firstfoot{ }
  \def\@journalName{}
  \def\@journalNameShort{}
  \def\@permissionCodeOne{}
  \def\@permissionCodeTwo{}
\def\formatline{ }
\def\permission{ }
\makeatother

\acmVolume{2}
\acmNumber{3}
\acmArticle{1}
\articleSeq{1}
\acmYear{2016}

\acmMonth{4}

\doi{}

\issn{1234-56789}

\usepackage{comment}
\usepackage{natbib}
\usepackage{graphicx}
\usepackage{textcomp}
\usepackage{xspace}
\usepackage{verbatim}
\usepackage{ulem}
\usepackage[rightcaption]{sidecap}
\usepackage{xcolor}
\newcommand*{\change}[1]{\textcolor{black}{#1}}
\newcommand{\todo}[1]{}
\newcommand{\remove}[1]{}
\newcommand{\refurl}[1]{\xspace({\small \url{#1}})}

\newcommand{\cendari}{CENDARI\xspace}

\title{The \cendari Infrastructure}
\author{
Nadia Boukhelifa  \affil{Télécom ParisTech \& CNRS LTCI}
Mike Bryant \affil{KCL}
Nataša Bulatović \affil{Max Planck Digital Library, Max Planck Society}
Ivan Čukić \affil{Faculty of Mathematics, University of Belgrade}
Jean-Daniel Fekete \affil{INRIA}
Milica Knežević \affil{Mathematical Institute of the Serbian Academy of Sciences and Arts}
Jörg Lehmann \affil{Universität Bern}
David Stuart \affil{KCL} %
Carsten Thiel \affil{Universität Göttingen, Niedersächsische Staats- und Universitätsbibliothek}}
\date{April 2016}

\begin{abstract}
The \cendari \emph{infrastructure} is a \change{research supporting platform}\remove{the technical result of a European project} designed to provide \remove{and support}tools for transnational historical research, focusing on two \remove{periods}\change{topics}:  \remove{the middle-ages}\change{Medieval culture} and \remove{World-War I}\change{World War I}. \remove{The visible tools are web-based, using modern web-oriented technologies}\change{It exposes to the end users modern web-based tools}\remove{ . They rely} \change{relying} on a sophisticated infrastructure to collect, enrich, annotate, and search through large document corpora.
Supporting researchers in their daily work is a novel concern for infrastructures\remove{,}\change{.}\remove{and we} \change{We} describe how we gathered requirements through multiple methods to understand the historians' needs and \remove{devise}\change{derive} an abstract workflow to support \change{them}.
We then \remove{describe}\change{outline} the tools we have built\remove{in more details}, tying their technical descriptions to the \remove{human} \change{user} requirements.  The main tools are the Note\remove{-} Taking Environment and its faceted search capabilities, the Data Integration platform including the Data API, supporting semantic enrichment through entity recognition, and the environment supporting the software development processes throughout the project \remove{development} to keep \change{ both technical partners and} researchers in the loop. 
The outcomes are technical \change{together with} \remove{, but also related to} new resources developed and gathered, and\remove{to} the research workflow that has been described and documented.

\end{abstract}



\acmformat{N. Boukhelifa, M. Bryant, N. Bulatović, I. Čukić, J-D. Fekete, M. Knežević, J. Lehmann, D. Stuart and C. Thiel: Cendari-Infrastructure}

\begin{document}
\begin{bottomstuff}
The research leading to these results has received funding from the European Union Seventh Framework Programme ([FP7/2007-2013] [FP7/2007-2011]) under grant agreement n\textdegree  284432. 
\end{bottomstuff}

\maketitle

\section{Introduction}

The \cendari infrastructure is the technical result of the \cendari project \cite{CENDARI2015}, a European Infrastructure project funded by the EU for 2012--2016. The infrastructure is designed to provide and support tools for historians \remove{as well as}\change{and} archivists. The \remove{visible }tools are web-based, using modern web-oriented technologies \change{and standards}. 
The \cendari infrastructure is innovative because it is designed to address multiple scenarios with two types of actors: researchers and cultural heritage institutions providing data. Both benefit from the platform, although the novelty\remove{lies more on the researcher's side} \change{concerns the researchers more}. 
To address researchers' needs, the platform provides a note-taking environment to perform the initial steps of historical research, such as gathering sources, writing summaries, elaborating ideas, planning \remove{, and}\change{or} transcribing\remove{, to name a few}. \remove{Also to support research, }\cendari integrates \change{online available resources initially}\remove{already available online and} curated by cultural heritage institutions with the explicit goal of integrating them into the infrastructure \change{to support the research process}. Researchers from the project visited these institutions and negotiated data sharing agreements. For archivists, an Archival Directory was \remove{integrated, which allows the establishment of archival descriptions and supports the international standard Encoded Archival Description (EAD)}\change{implemented, which allows the description of material based on the international Encoded Archival Description (EAD) standard \refurl{http://www.loc.gov/ead/index.html}}.
The resulting infrastructure \remove{provides }not only \change{provides} access to more than \change{800,000}\remove{of these descriptions of sources} \change{archival and historical sources}, but integrates them into a collection of tools and services developed by the project\remove{,} as a digital resource for supporting historians in their daily work\remove{flow}. 

The researchers' primary entry point into \cendari is via the \textit{Note-Taking Environment (NTE)}. It \change{enables}\remove{allows the} curation of notes and documents \remove{related to}\change{prepared by researchers within} various individual research projects. Each project can be shared among colleagues and published once finished. These notes and documents can be linked to the data existing in the \textit{\cendari Repository}\remove{ and knowledge base}. This comprises both the linking of entities against standard references such as DBPedia and connection to \remove{the}archival descriptions.
A faceted search \change{feature} is part of the NTE and provides access to all \remove{of \cendari's data. The search also}\change{data, and in addition} connects to the TRAME \remove{web client}\change{service~\protect\cite{TRAME}} for extended search in \change{distributed} medieval \change{databases}\remove{knowledge bases}.
\remove{The data is kept in the Repository based on the CKAN data management system which, in addition, provides a user interface to access the collected data.}\change{The Repository is based on the CKAN software\change{\refurl{http://ckan.org/}}, manages the data, and provides a browser-based user interface to access it.}

To\remove{enlarge its data set of archival descriptions} \change{support the creation of curated directories of institutions holding relevant material for both research domains, and to raise awareness about the ``hidden'' archives that do not have \change{a }digital presence or are less known but relevant, we integrated} the AtoM software\change{\refurl{{https://www.accesstomemory.org}}}\remove{is also integrated}, enabling historians and archivists to add new and enrich existing archival descriptions \change{in a standardised way, following the EAD}.

Once data is collected, an internal component\change{,} called the \textit{Litef Conductor (Litef)} processes it for further semantic enrichment. It \change{then} sends the text extracted from the documents \remove{'s contents}to the\remove{text search engine} ElasticSearch\refurl{{http://www.elastic.co/products/elasticsearch/}} search engine, \change{invokes}\remove{applies named entity recognition through the use of} the high-quality \textit{Named Entity Recognition and Disambiguation Service (NERD)} \remove{web-service}~\cite{NERD},
and generates semantic \remove{web triples}\change{data} inferred from several document formats such as archival descriptions or XML encoded texts.

The connection to the semantic web is further extended through ontologies developed specifically for the intended use cases and connected through the\remove{semantic store} \change{\textit{Knowledge Base (KB)}}.
\change{Researchers can upload their own ontologies}\remove{The ontologies can be viewed and a researcher's own ontology can be uploaded} to the \change{Repository} through \change{the \emph{Ontology Uploader}}\remove{a dedicated} tool.
To explore the semantic data collected\remove{ by \cendari}, the \textit{Pineapple Resource Browser} 
provides \remove{an intuitive}\change{a} search and browse web interface. 

To summarise, the \cendari technical and scientific contributions are:
\begin{itemize}
    \item The use of participatory design sessions as a method to collect users' needs,  
    \item The design of the \remove{\cendari }infrastructure as a research support tool,
    \item The deep integration of multiple resources through a \textit{data integration and enrichment process} to further support the researchers' workflow.
\end{itemize}

\noindent \remove{The} \cendari \remove{project}has generated several benefits and outcomes:
\begin{itemize}
    \item An integrated system to conduct the early stages of historical research,
    \item A catalogue of hidden archives in Europe (more than \change{5,000} 
       \change{curated institutional and} archival descriptions \remove{manually created} with special attention given to archives without own digital presence),
    \item A rich set of Archival Research Guides to help and guide historians by providing transnational access to information\change{~\cite{cendariwhitebook2015}},
    \item Integrated access to heterogeneous data from multiple data sources and providers,
    \item\change{A rich set of resources accessible and searchable from the web.}
\end{itemize}

\section{Related Work}

\cendari is related to several other \change{Digital Humanities} (DH) projects \remove{(some currently active) }which \remove{attempt to}\change{tackle aspects} of the humanities research workflow from either a generalist or topic-specific perspective. 
The project is directly connected to \change{the \textit{Digital Research Infrastructure for the Arts and Humanities(DARIAH)}\refurl{http://dariah.eu/}}\change{, harmonizing national DH initiatives across Europe,}\remove{the Digital Research Infrastructure for the Arts and Humanities.}\remove{, an ESFRI project established as ERIC in 2014}
\change{with a}\remove{ The} goal\remove{ of DARIAH,which is composed of several national initiatives across Europe, is } to build and sustain an infrastructure that supports and enables the research of humanities scholars in the digital age. From its inception, \cendari was designed to \change{rely on and} ultimately hand over the resulting infrastructure to DARIAH for sustainability \change{and reuse}.

The \textit{European Holocaust Research Infrastructure (EHRI)} project aims to assist the process of conducting transnational and comparative studies \change{of}\remove{into} the Holocaust by bringing together in an online portal \refurl{https://portal.ehri-project.eu/}\remove{descriptions of} archival material across a geographically dispersed and fragmented archival landscape\cite{blanke2013integrating}. \remove{In this, }\change{Its} principle goal is related to data collection, integration, and curation.

The \change{\textit{TextGrid Virtuelle Forschungsumgebung für Geisteswissenschaften} \refurl{https://textgrid.de/}}\remove{ran from 2006 to 2015 under funding by the German Federal Ministry of Education and Research (BMBF).
The project} developed a virtual research environment targeted specifically at the humanities and textual research. \remove{With its large existing user base and adoption within further research projects, TextGrid has become a key technology and platform in the Digital Humanities.}
\change{Its open source components are implemented as a desktop client to work with TEI encoded documents and to create and publish scholarly editions.}
\remove{Therefore,}\change{Following} the \remove{project's end, the} \change{end of the project the} infrastructure components and technology were integrated into \remove{the}DARIAH. \remove{research infrastructure.}

To understand the needs of historians in the context of transnational history, we have used two key methods: the development of use cases and participatory design sessions. Participatory design \remove{is an approach to design}\change{has been} adopted by many disciplines where stakeholders cooperate to ensure that the final product meets their needs\remove{ and is usable}.  For interactive software design, which is the focus of this work, the aim is to benefit from different expertise: designers know about the technology and users know their workflow and its context. In \remove{the same context}\change{a similar vein}, Muller and Druin~\cite{muller:2002} describe participatory design as belonging to the \textit{in-between} domain of end-users and technology developers which is characterised by reciprocal learning and the creation of new ideas through negotiation, co-creation\change{,} and polyvocal dialogues across and through differences.

Until recently, participatory design was not a common practice in humanities projects. Claire Warwick~\cite{warwick:2012} provides explanations for this paucity: \textit{
It was often assumed that the resources created in digital humanities would be used by humanities scholars\remove{, who were not technically gifted}... there was little point asking them what they needed, because they would not know, or their opinion about how a resource functioned, because they would not care. It was also assumed that technical experts were the people who knew what digital resources should look like\remove{, what they should do and how they should work}... their opinion was the one that counted, since they understood the details of programming, databases, XML and website building. The plan, then, was to provide good resources for users, tell them what to do and wait for them to adopt digital humanities methods''}. Digital humanities has since then changed perceptions and we can now see a number of projects adopting participatory design to learn about users and their requirements \cite{mattern:2015,wessels:2015,visconti:2016}.
\section{Design and Development Process}\label{sec:design}
\remove{Supporting}\change{Assisting} historians and archivists \remove{for}\change{in} their research \remove{in}\change{on} transnational history is \change{an} important but abstract \change{goal}. To design relevant tools, \cendari \remove{project} first had to understand \remove{the}related needs and requirements of historians and archivists.  It turned out that the work process of these researchers is not well described in books or research articles so the initial design phase of the \remove{\cendari}project consisted in obtaining information about the needs and \change{practised} research processes of our target researchers. 

\begin{figure}

  \centering
  \includegraphics[width=0.8\linewidth]{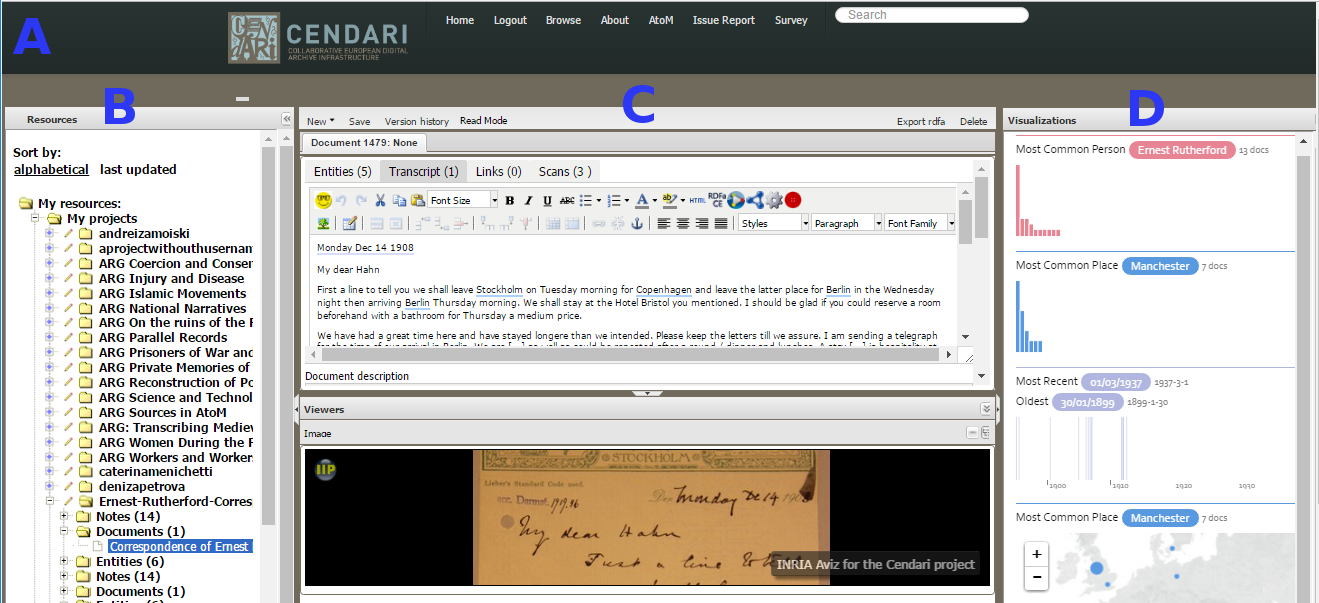}
  \caption{\label{fig:NTE}
           The Integrated \remove{Note-Taking Environment (NTE)}\change{NTE} for Historians: (A) the search and browse panel, (B) the resources
panel, (C) the central panel dedicated to editing, tagging and resolution, and (D) the visualisation panel.}
\end{figure}

\subsection{Use Case}

Let's start by an example of a historian's workflow\change{, acting}\remove{and } as a use case for historical research carried out in \remove{the}\cendari.\remove{ infrastructure.}
It has been developed alongside \remove{with }the \textit{Archival Research Guide(ARG) on Science and Technology}\remove{during and after the First World War} written by one of the historians in the \change{project}\remove{\cendari team}. This \change{ARG}\remove{research guide} examines the shift within the transnational scientific network during and after the First World War \change{(WWI)}, which resulted in the isolation of the German Empire within the international scientific community. The research question is whether this shift can also be found in the correspondence of individual scientists. There was an intense international exchange between physicists before the war, and a particular prominent example is the case of the British \change{scientist} Ernest Rutherford, later Nobel Price winner and president of the British Royal Society. 

As a first step, a new project with the name ``Ernest Rutherford Correspondence'' was created in the NTE (see Fig.~\ref{fig:NTE}). Afterwards \remove{the} available resources were searched through the faceted search, integrated in the NTE, and \remove{the}\change{selected} results \change{(consisting of descriptions of archival material)} were saved into a note.\remove{The selected results consist of descriptions of archival material.}

The next step \remove{consisted of}\change{was} a visit to an archive in Berlin, which \change{holds} the biggest collection of Rutherford’s correspondence in Germany. Several notes were taken there and added to the project space in the NTE. Photographs of the letters were taken, uploaded to the private research space and later transcribed in the working space, using the provided image viewer. 

The transcription of each document was processed with the \remove{Named Entity Recognition and Disambiguation Service (}NERD\remove{)} \change{service, integrated within the NTE}. This service produces many results; since the visualisation of these entities cannot always be done automatically\remove{,} the user might have to resolve entities manually, for example by selecting the appropriate place or person from the list of alternatives provided in the resources panel of the NTE. The visualisations on the panel to the right \remove{hand}show Ernest Rutherford’s correspondents in a histogram, on a map the places from where he wrote this letters (such as Montreal and Manchester, where he lived from 1907 onwards), and the timeline shows how Rutherford’s German correspondents did not receive any letters from him after 1914. In this way, the research hypothesis - abrupt ending of exchanges between German and Anglo-Saxon scientists from the beginning of the First World War onwards - is being substantiated.

From the list of resources available in archives and libraries it can be discovered that Cambridge University Library holds the Correspondence and Letters of Ernest Rutherford and thus the most important and comprehensive part of his heritage. If \change{the researcher} is not able to visit Cambridge \remove{by himself, f.ex.}\change{for example,} because of lack of financial support, the researcher can ask a colleague abroad to contribute to the \change{endeavour} by visiting Cambridge and checking the correspondence in the archive. In order to enable a researcher abroad to contribute to the research project, it can be shared with him by ticking a box in the project edit page. That does not mean that the project is \change{publicly visible}\remove{in the public domain}, as all content is inherently private in the NTE. The project is simply being shared with one colleague based in another country. \remove{But this way}\change{In this manner}, collaborative and transnational research becomes possible. Once the material in Cambridge has been explored and partially described, more complex interpretations and the deeper layers of the research question can be pursued. \remove{In the given case, this would be f.ex.}\change{In this case, for example,} the correspondence and interaction between Ernest Rutherford and Otto Hahn in the early 1930s, and Rutherford's influence on Hahn's discovery of the nuclear fission.

\subsection{\remove{\cendari }Participatory Design Workshops}

There are many participatory design methods including sitings, workshops, stories, photography, dramas and games~\cite{muller:2002}.\remove{For \cendari,} \change{We} were inspired by low-fidelity prototyping methods~\cite{beaudouin-Lafon:2002} because they provide concrete artifacts that serve as an effective medium for communication between users and designers. In particular, to explore the \cendari virtual research environment design space, we used brainstorming and video prototyping. 
Together, these two techniques can help explore new ideas and simulate or experience new technology.

For brainstorming, \remove{usually a group of three to seven people works best. The group}\change{a group of people, ideally between three and seven in number,} is given a topic and a limited amount of time. Brainstorming has two phases: an idea generation phase and an evaluation phase. The aim of the first stage is to generate as many ideas as possible. Here quantity is more important than quality. In the second stage, ideas are reflected upon and only a few \remove{are} selected for further work (e.g. video prototyping). The selection criteria could be a group vote where each person picks their favourite three ideas.

Video prototyping is a collaborative design activity that involves demonstrating ideas for interaction with a system in front of a video camera. Instead of describing the idea in words, participants demonstrate what it would be like to interact with the system. The goal is to be quick and to capture the important features of the system that participants want to be implemented. A video prototype is like a storyboard: participants describe the whole task including the user's motivation and context at the outset and the success of the task at the end. In this way, the video prototype is a useful way to determine the minimum viable system to achieve the task successfully.

We organised three participatory design sessions~\cite{boukhelifa2015} with three different user groups: WWI historians, medievalists, and archivists and librarians. The aim of these sessions was to understand how different user groups would want to search, browse, and visualise (if at all) information from archival research. In total there were 49 participants (14 in the first, 15 in the second and 20 in the third). Each session was run as a one-day workshop and was divided into two parts. The morning began with presentations of existing interfaces for access and visualisation. In order to brainstorm productively in the afternoon, participants needed to have a clear idea of the technical possibilities currently available. In the afternoon, participants divided into $3-5$ groups of four and brainstormed ideas for searching, browsing, and visualisation functions, and then they created paper and video prototypes for their top three ideas. There were 30 video prototypes in total, each consisting of a brief (30 seconds to 4 minutes) mock-up and demonstration of a key function. Everyone then met to watch and discuss the videos.

\begin{figure}
  \centering
  \includegraphics[width=0.7\linewidth]{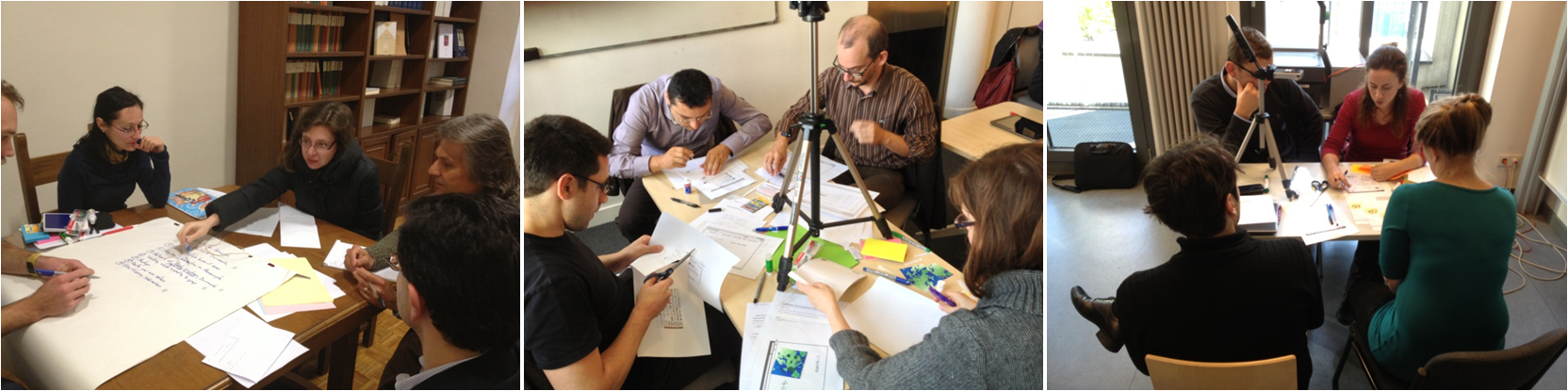}
  \caption{\label{fig:PDSs}
           The three participatory design sessions held with WWI historians, librarians and medievalists.}
\end{figure}

\noindent\textbf{Findings: }
These participatory workshops served as an initial communication link between the researchers and the technical developers and they were an opportunity for researchers to reflect on their processes and tasks, both those they perform using current resources and tools and those they would like to be able to perform. Even though there were different groups of users involved in these workshops, \remove{there were} common themes \remove{that} emerged from the discussions and the prototypes. The outcomes of the participatory sessions were three-fold: a set of functional requirements common to all user groups, high-level recommendations to the \remove{\cendari} project and a detailed description of historians workflow (section ~\ref{sec:workflow}).

In terms of functional requirements, participants expressed a need for: \textbf{networking} facilities (e.g. to share resources or to ask for help); robust \textbf{search} interfaces (e.g. search for different types of documents and entities or by themes such as language, period or concept); versatile \textbf{note-taking} tools that take into consideration paper-based and digital workflows and support \emph{transcription} of documents and \emph{annotations}; and interactive \textbf{visualisations} to conceptualise information in ways that are difficult in text-based forms.

\remove{Besides the functional requirements, we were able to conclude the}\change{The} participatory design workshops \change{were concluded} with the following high-level recommendations \change{to \cendari}: \textbf{(i)} \remove{\cendari needs }to combine existing workflows with new digital methods in ways that save researchers' time. In particular, notes can be augmented over time\change{,} and researchers willingness to share them might depend on the note-taking stage and their motivation for sharing\remove{. \cendari needs to take into consideration the dynamic nature of note-taking and changing user requirements}; 
\textbf{(ii)} all researchers have data that they do not use after publication or end of their projects. If \remove{\cendari}\change{this project} can offer benefits for making this data available with proper citations, such \change{an} initiative could encourage researchers to release their data. This could change working practises and bring more transparency to historical research. Indeed, linking the management of research data to publication and presenting tangible benefits for researchers are important \remove{to attract}\change{factors in attracting} contributors; and \textbf{(iii)} \remove{Workshop participants suggested that \cendari should }\change{to} work closely with researchers to develop testbeds early in the project rather than seek feedback at the end of software development. Researchers who are currently working on a project are likely to have useful data and an interest in sharing it. These projects could be implemented as use cases demonstrating the utility of \change{our}\remove{the \cendari} virtual research environment. 

In order to create a technical system that can adequately support these functional requirements and recommendations, information collected from the participatory workshops was translated into precise descriptions of functions, which were then evaluated by technical experts and used as the basis for technical development. As part of the followup from these workshops, it was decided to translate important functions demonstrated in the video prototypes into a written format in the form of use cases. An example of such use cases is described \change{at the beginning of} section~\ref{sec:design}.


\subsection{Historian Research Workflow}\label{sec:workflow}

Through the discussions and exchanges gathered during the participatory design sessions and standard references~\cite{historiography}, \change{\cendari}  technical partners \remove{of the \cendari project}identified a \remove{the following} workflow that seems to match the work of a broad range of historians in the early phases of their research, although we do not claim that every historian follows it. We summarise it here to refer to it later \remove{in relation with}\change{concerning} the tools and mechanisms that the infrastructure supports.

\begin{enumerate}
    \item\label{wf:preparation} \emph{Research preparation}: all historians start a new project by gathering questions, hypotheses, and possible ways of answering or validating them. This phase is usually informal and carried-out using a note-book, either paper-based or computer based. During the participatory design sessions, researchers complained that the notes they take are hardly organised, often spread over many computer files or notebook pages.
    \item\label{wf:selection} \emph{Sources selection}: to answer the questions and validate the hypotheses, researchers gather books, articles, web-accessible resources, and make a list of primary and secondary sources to read. This stage is repeated each time new information is collected.
    \item\label{wf:planning} \emph{Planning of visits to Archives and Libraries}: relevant sources are often accessible only from specific archives or libraries, because the sources are unique; even the catalogues \remove{or metadata }are not always accessible online, or not precise enough to answer questions. Physical visits to the institutions are then necessary\remove{, usually from the closest to the farthest, weighted by the chance to find relevant documents. Often not enough resources for research
    travels are available}.
    \item\label{wf:visits} \emph{Archive and Library visit}: working at archives or libraries involves note-taking, transcribing, collecting scans and photos of documents for later exploitation. Sometimes, even more work is involved when archive boxes have never been opened before and some exploration and sorting is needed, hopefully (but not always) improving the catalogues.
    \item\label{wf:notes} \emph{Taking Notes}: during their visits in archives and libraries, historians take notes or \remove{write down annotations to the primary sources inspected in situ}\change{annotate copies of archival records}. These notes and annotations generally follow the main research interests of the \remove{scientists}\change{researchers}, but also contain side glances to related topics or possible research areas. Most of the time they can be understood as in-depth descriptions of the sources and thus as a complement to the metadata available in finding aids.
    \item\label{wf:transcription} \emph{Transcription}: primary sources consulted are often \emph{transcribed}, either partially or exhaustively to facilitate their reading, but also to enhance their searchability. These transcriptions serve as the basis for literal citations in publications.
    \item \emph{Research refinement and annotation}: from the information gathered, some questions are answered, some hypotheses are validated or invalidated, but new questions arise as well as new hypotheses.  In particular, this list of questions comes repeatedly with regard to the following facts:
        (a) who are the \emph{persons} \remove{involved}\change{mentioned} in the documents. Some are well known, others are not and yet are frequently mentioned. Understanding who the persons are and why they \remove{are}\change{were} involved is a recurring question in \remove{this repeated stage}\change{the research process}.
        (b) where are the \emph{places} mentioned? Finding them on a map\remove{, understanding the geography} is also a recurring \remove{question}\change{issue}.
        (c) what are the \emph{organisations} mentioned in the documents, and their relationship with the persons and places.
        (d) clarifying the temporal order of events is also essential and dates often appear \remove{in unclear formulations, }with a high level of uncertainty.
    \item\label{wf:struct} \emph{Knowledge organisation and structuring}: after enough facts have been gathered, historians try to organise them in high-level structures, with causalities, interconnections, or goals related to persons and organisations. This phase consists also in taking notes, but also in referring to other notes, transcriptions, and by organising the chunks of information from previous notes in multiple ways, not related to the documents where they come from but rather from higher-level classifications or structures.
    \item\label{wf:refinement} \emph{Refinement and writing:} at some point, the information gathered and organised is sufficient for writing an article or a book. All the bibliographic references are gathered, as well as the list of documents consulted in archives and libraries, to be \remove{included}\change{referenced} in a formal publication. 
    \item\label{wf:expansion} \emph{Continuation and Expansion:} often a research work is reused later either as a continuation of the initial research, or with variations reusing some of the structures (other places, other organisations, other times).
    \item\label{wf:collaboration} \emph{Collaboration support}: while collaborations are possible and realised in the physical world, sharing of gathered material is limited due to the difficulty of copying every physical information between collaborators. In contrast, a web-based setting allows the sharing of all the resources related to a research project. Historians and archivists have expressed the desire, during all the participatory design sessions, to experiment digital-based collaborations.
\end{enumerate}


Archivists \remove{are not following}\change{do not follow} the same workflow but they share some of the concerns, in particular \change{the need} to identify persons, places, organisations, and temporal order\remove{. T}\change{, since t}his information is essential to their cataloguing activity. They also have \change{a} higher-level \remove{questions useful to make}\change{engagement with the material that is useful in making} sense of institutional logic or just understanding the contents and provenance of particular boxes and collections. 
We note that the workflow described above is non-linear as reported in other studies (e.g.~\cite{mattern:2015}). Researchers can at any stage acquire new information\change{, reorganise}\remove{ that makes them reorganise} their data, refine their hypotheses or even plan new archive or library visits. 

The \cendari platform has been built to support this workflow with web-based tools, and to augment it with collaboration\remove{s}, sharing, and faceted search through the gathered and shared documents. \change{Main}\remove{The central} functional requirements can be summarised as: taking notes, transcribe, annotate, search, visualise, and collaborate.

\subsection{Iterative Approach and Support Technologies}\label{subsec:iterativeapproach}
The software development in the project was carried out by adopting agile development methods. \remove{In particular}\change{The} software components were developed in short iterative release cycles and direct user feedback was encouraged and incorporated.
\remove{As the}\change{The software is}\remove{ software created was} released as open source \change{and the code is}\remove{, the code was} hosted on and maintained through GitHub~\cite{Cendari:dev}.
Where possible\remove{,} \change{\cendari}\remove{ project} used the existing solutions \remove{from}\change{provided by} DARIAH\change{, such as the JIRA\refurl{{https://www.atlassian.com/software/jira}} ticketing system}.
\remove{Thus when the}\change{Researchers who contacted relevant}\remove{started contacting cultural heritage} institutions for inclusion of their holdings into the platform \remove{, the JIRA\refurl{{https://www.atlassian.com/software/jira}} ticketing system was chosen}\change{used it} to support and document the workflow from \change{the} first contact through negotiations and agreements for data sharing, to the final ingestion. 
Following the positive experience of historians using the tool\remove{ for this workflow}, JIRA remained the obvious choice for bug and issue tracking during the software development.
By tightly integrating the applications with the DARIAH development platform and establishing close communication between historians and developers, all parties involved in the development process were able to engage in direct and problem oriented development cycles.
Ultimately, the release cycles were reduced to one week \change{and included}\remove{going from} dedicated test and feedback sessions \change{as well as}\remove{to} development focusing on small numbers of issues that were identified and prioritised collaboratively in the group.

One of the decisions that enabled these rapid cycles was the adoption of a DevOps model to \remove{managing and developing}\change{manage and develop} the \cendari infrastructure.
By applying methods of agile development to system administration and simultaneously combining the development and management of the applications and infrastructure, as discussed in 
e.g.~\cite{Kim:PhoenixProject},
the effort and amount of required work\remove{needed to go} from code commit to deployment was dramatically reduced. This was achieved by implementing automated code build and deployment using \remove{a}\change{the} Jenkins~CI\refurl{{https://jenkins.io/}} platform and Puppet\refurl{{https://puppet.com/}} configuration management on \remove{both} \change{dedicated} staging and production servers\remove{, the infrastructure developed in parallel to the software it supported}.


\section{The \cendari Infrastructure}

The \cendari infrastructure \change{technically}\remove{is meant to provide tools and resources for historians and archivists in order to enable them to conduct transnational historical research.} 
\remove{It}\change{combines \emph{integration} of existing components and tools, \emph{extension} of other components and tools and \emph{tailored development} of the missing pieces.}
The overall goal was to offer a great user experience to the researchers, part of which was already implemented by existing tools, while avoiding development of an infrastructure from scratch\remove{ and keeping it under control}.

Finding a set of tools that can be either developed or combined to form an integrated environment was \remove{sometimes an underestimated}\change{a} challenge. We realised that \change{in order to address the user requirements} \remove{we needed}\change{it was necessary} to provide \remove{more than a single tool}\change{multiple tools} to support several stages of the researchers' workflow\change{,} and the production of research outputs and formats, as no \change{single} tool could offer all the required features. 
\change{The challenging part was to select and decide upon}\remove{with} a limited number of components\change{, technologies and tools}\remove{and technologies} \change{which users could use intuitively and without extensive training}.

\begin{figure}\centering
\includegraphics[trim={0 2cm 0 0},clip,width=0.8\textwidth]{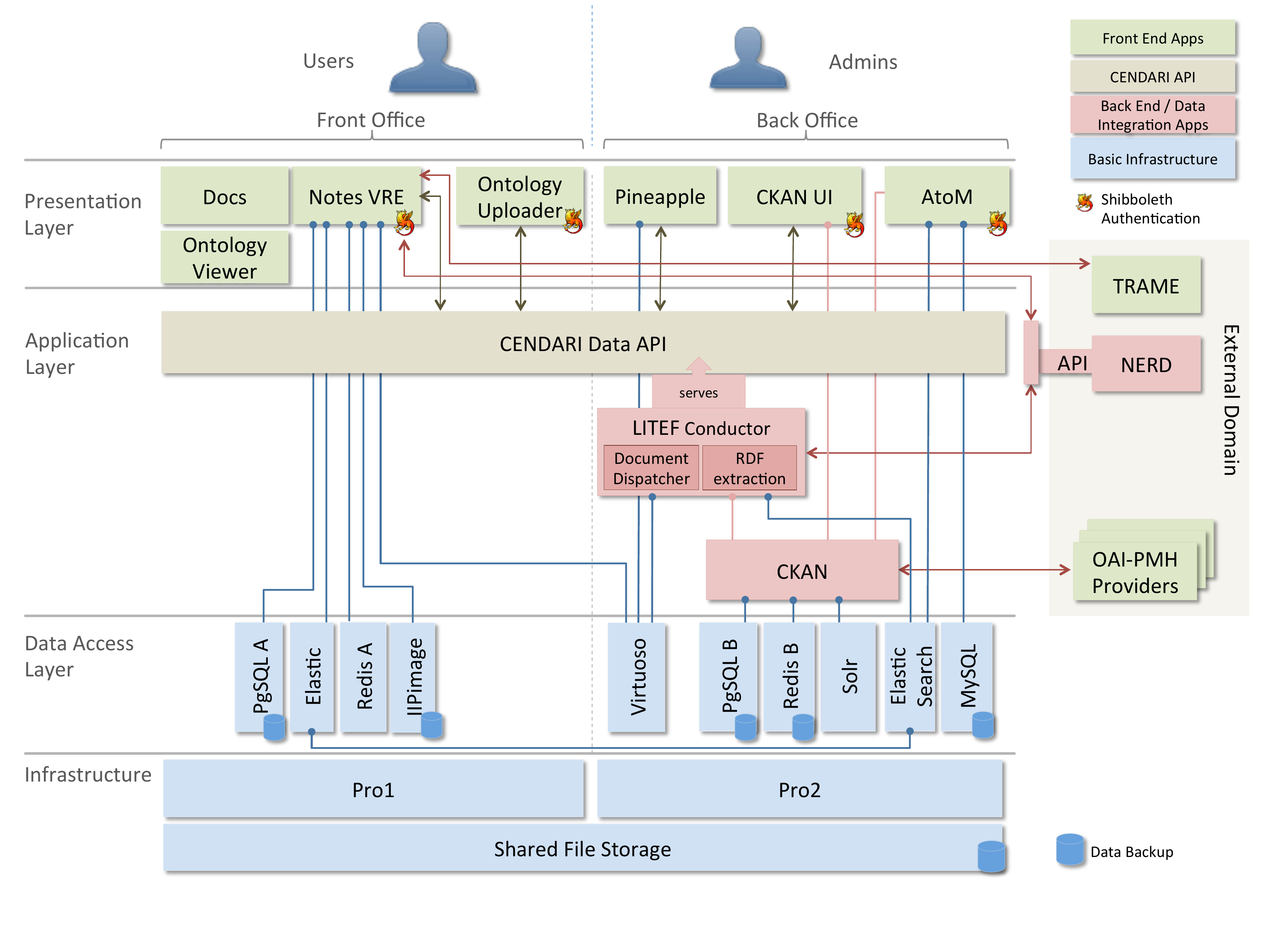}
\caption{\cendari Infrastructure Model, originally by Stefan Buddenbohm}
\label{fig:infrastructuremodel}
\end{figure}

To implement this infrastructure, a modular approach was undertaken.
In using existing tools and services, we were able to offer many tools for some of the major features.
At the same time, several important parts of the infrastructure (see Fig.~\ref{fig:infrastructuremodel}) were developed \change{especially for \cendari.}\remove{with a particular use in the \cendari infrastructure in mind}

At the data \change{persistence} level\remove{,} we \remove{use} \change{ take a} polyglot \change{approach}\remove{ persistence through four main engines}: relational databases such as PostgreSQL\refurl{{http://www.postgresql.org}} and MySQL\refurl{{https://www.mysql.com/}} for most web-based tools, ElasticSearch for our faceted search, and the Virtuoso{\refurl{http://virtuoso.openlinksw.com} triple store to manage \remove{\cendari- generated semantic-web data, plus other}\change{ generated semantic data and and triples created by the NTE}\remove{web-based tools}.

In the application layer, the central data exchange and processing is carried out by the \emph{Data API}, a component implemented within the project (see Section~\ref{sec:dataapi}).
The data is stored in \change{\emph{the Repository}, based on CKAN}\remove{a CKAN repository so it} \change{and} can be accessed through \remove{the web}\change{a web browser}. \change{Within the project, several extensions to \change{CKAN}\remove{the CKAN software} were developed to support the data harvesting and user authentication mechanisms.}
\remove{By using the well-known and widely used CKAN repository, t}\change{T}he development \remove{on}\change{of} the \emph{Data API} and the \emph{Litef Conductor} \remove{indexing service} focused on providing \remove{the} specific \change{data services}\remove{features} required for the project.

Several user applications in \cendari support the historian's workflows (see ~\ref{sec:workflow}): the \emph{\remove{Note-Taking Environment (NTE)}\change{NTE}}, the \emph{Archival Directory}~\cite{ADT} and the \emph{Pineapple Resource Browser Tool}\refurl{https://resources.cendari.dariah.eu/}. 

The \change{NTE}\remove{Note-Taking Environment (NTE)}, described in detail in Section~\ref{sec:nte}, combines access to the faceted search and the repository data with 
individual research data. The NTE is an adaptation and extension of the EditorsNotes system\refurl{{http://editorsnotes.org}}.
The Archival \change{Directory}\remove{ Description Tool}, \change{based on the AtoM software}, is used for manual creation and curation of archival descriptions.  It has a strong transnational focus and includes ``hidden'' archives and institutions, ``little known or rarely used by researchers''~\cite{ADT}. At present it offers \change{information about more than 5000 institutional and archival descriptions curated by \cendari researchers}}.
\remove{The Resource Browser Tool (}Pineapple\remove{)} provides free-text search and faceted browsing through our \emph{Knowledge Base \change{(KB)}}, containing resources and ontologies from both domains \change{(WWI and Medieval)}. \remove{In the background,}\change{Technically,} Pineapple is a SPARQL client, sending predefined parametrised queries to Virtuoso to extract and render information about available resources, related entities such as people, places, events or organisations, or resources which share same entity mentions from different data sources. These resources are generated by the semantic processing services\remove{of \cendari } (see \ref{sec:semanticext}) or are integrated from the Medieval knowledge base \change{TRAME (see below)}. \remove{Besides the user interface }Pineapple \change{provides a web-based user interface, and} uses content negotiation\refurl{{https://www.w3.org/Protocols/rfc2616/rfc2616-sec12.html}} \remove{and acts}\change{to provide} \remove{as }a REST-based (non-SPARQL) interface to the \remove{\cendari}KB, delivering JSON formatted data.
%
\change{Advanced} and ontology-savvy users can \change{use}\remove{access} the \emph{Ontology \remove{visualisation}\change{Viewer}} tool or upload own ontologies via the \emph{Ontology Uploader}\remove{tool}.

All \cendari services authenticate users through the DARIAH~AAI~\refurl{https://wiki.de.dariah.eu/x/JQurAg}, a federated authentication solution based on Shibboleth\refurl{{https://shibboleth.net}}, for which dedicated plugins for AtoM and CKAN were also created.
The DARIAH~AAI handles the verification of users \remove{ and their involvement with the research community as well as}\change{and} the creation of user accounts including support for password management. \change{From an end user's perspective, this is a Single-Sign-On experience, where \cendari components are visible as one of many DARIAH services.}\remove{
This enables a clear integration of \cendari as a DARIAH service from the user's perspective as well as a Single-Sign-On experience between \remove{the}\cendari components and any other DARIAH service.}
While \remove{the}Shibboleth \remove{integration} is used \change{to authenticate users}\remove{for authentication of users}, access permissions to the individual resources within the \cendari infrastructure are handled by the Repository.

In addition to the main applications listed above, \remove{several}\change{two other components were} developed\change{/adjusted:}\remove{and hosted by external partners were re-used.}
\remove{The}\change{the} \emph{NERD service} \cite{NERD} and the\remove{medieval} TRAME application \change{to search for distributed medieval resources} \cite{TRAME}\change{. These} are hosted and maintained by the developing partner institutions and provided through defined and documented interfaces for \cendari and other third parties.

The setup of the main applications was implemented using configuration management \remove{following the defined layout of}\change{as shown in} Fig.~\ref{fig:infrastructuremodel}\remove{.
It}\change{, which} provides a high-level overview of the infrastructure design, distinguishing between the internal Application and Data Layers and the user facing Presentation layer vertically.
The model also shows the distinction between the Front Office and Back Office, split over two machines, separating the common user applications from those components used by power users only.
During the development of the infrastructure, two instances of both servers were used to allow for integration tests between components in a dedicated staging environment before deployment to the production servers.

The modular approach to designing this infrastructure is exemplified by the role of the Data API as \change{a} central \remove{application}\change{component} for data exchange between the user facing applications and the Repository, including user authorisation on the resource level.
All user-facing applications communicate trough the Data API. Thus by adopting the API any underlying repository solution could be used.

\change{\cendari}\remove{The \cendari project} was designed from the start \remove{to build}\change{as} an infrastructure that can ultimately be sustained \change{and reused} through DARIAH.
To achieve this, we used\remove{both the existing} tools and solutions offered by DARIAH \change{or other parties,} and integrated the services into \remove{their}\change{the} existing ecosystem where possible, such as the development platform, the AAI\change{, CKAN, AtoM}.
Additionally, the technical design of the infrastructure was aligned with the efforts undertaken \change{in parallel}\remove{concurrently} to migrate the TextGrid infrastructure into the DARIAH ecosystem.

\section{Note-Taking Environment}\label{sec:nte}


Most historians collect notes in files using standard text editors. The files are sometimes well sorted in folders, but most of the time \change{the files are} difficult to find due to the loose organisation of early research projects.  The information related to the notes are also scattered in multiple locations.  Sometimes, historians take pictures at archives or libraries; these pictures end-up stored in whatever location their camera or system stores them. The file names associated with these pictures are also unrelated to the notes or research projects.  Even if some historians are well organised, it takes them considerable time and discipline to organise their virtual working space on their own computer.  And even with the strongest discipline, it is almost impossible to link the multiple documents together, connecting notes to pictures, scans, PDF files, spreadsheets, or other kinds of documents they use.  With all these problems in mind, and to facilitate the collaboration between researchers, we have designed and implemented the NTE in tight integration with the whole \cendari infrastructure.
The NTE implements the historian's workflow described in Section~\ref{sec:workflow}.\remove{, in particular the stages \ref{wf:notes}, \ref{wf:transcription}, \ref{wf:struct}, \ref{wf:refinement}, \ref{wf:expansion}, and \ref{wf:collaboration}.}

\subsection{Overview}

The NTE is designed to manage \emph{documents} and \emph{notes} gathered for a \emph{project}.  Typically, a project is a thesis, or a journal article, but more generally, it is a container for gathering and linking information, refining it, collaborate, and prepare publications. The final publication or production is not done inside the NTE\remove{,} \change{since} there are already many editing environments to perform that task.

The main user interface of the NTE has three main panels \change{coordinated using brushing and linking~\cite{brushing}}: \remove{besides}\change{the search} and browse panel (Fig.~\ref{fig:NTE}(A)); a library where the user can manage projects and browse allocated resources (Fig.~\ref{fig:NTE}(B)); a central space for editing, linking and tagging resources (Fig.~\ref{fig:NTE}(C)); and a visualisation space for showing trends and relationships in the data (Fig. \ref{fig:NTE}(D)). 

\noindent\textbf{The resources panel}: resources are organised per project into three main folders \remove{that}\change{corresponding} roughly to the way historians organise their material on their machines. The notes folder contains files, each file is a note describing archival material related to a project. The user can select a note and its content is shown in the central panel. The user can edit the note, tag words to create entities such as event, organisation, person, place and tag. \change{She}\remove{Users} can add references to documents\remove{. The documents themselves} \change{which} can be letters, newspaper articles, contracts or any text that acts as evidence for an observation or a statement in a note. Documents can contain named entities, a transcript, references to other documents and resources, as well as scanned images, which are displayed in the high resolution web-based image viewer at the bottom of the central panel. 

\noindent\textbf{The central panel:} \remove{it} acts as a viewing space for any type of resource and \remove{thus}mimics the function of a working desk of a historian. This is \remove{also}where entity tagging and resolution takes place. \change{The user} may not be entirely clear about the true identity of an entity, for instance, in the case of a city name that exists in different countries. \remove{The user has the option to}\change{She has an option to} manually resolve \change{it}\remove{, when further information becomes available}, by assigning a unique resource identifier URI (e.g. to a unique Wikipedia entry).

\noindent\textbf{The visualisation panel:} \remove{it} provides useful charts to show an overview of entities, distributions, frequencies and outliers in the resources of the project\change{, and}\remove{. There is also} a map which shows the location of any \emph{place} entity.
\remove{The three panels are coordinated using brushing and linking~\cite{brushing}. In terms of privacy settings, by default notes are private and entities are public, but users can change these permissions.} 

In summary, \change{the} NTE supports the following features: (1) \emph{Editing and annotation}: through a rich set of editing, formatting and tagging options provided by RDFaCE\refurl{{http://rdface.aksw.org/}} (2) \emph{Faceted Search}: \remove{a faceted search service} for thematic \change{search and} access to resources; (3) \emph{Visualisation}:  showing histograms of three entity types (names, places and events) \change{and}\remove{as well as} a geographical map with support for aggregation \change{(other visualisations could be easily integrated)}; (4) \emph{Automatic Entity Recognition}: in the form of an integrated \change{NERD} service in the editor (5) \emph{Interaction}: the visualisations support selection, highlight and pan\&zoom for the map.  Brushing and linking is implemented with one flow direction for consistency from left to right. This is to support the workflow in the NTE: select resource, view\&update, then tag and visualise. Note that \change{in addition} referencing, collaboration and privacy setting are available in the NTE. \change{In terms of privacy settings, by default notes are private and entities are public, but users can change these permissions.}

\subsection{Technologies}\label{sec:nte-technologies}


\change{The NTE implements}\remove{ We use} a client-server architecture\remove{; on the client side we rely on modern web technologies}\change{, the client side relying on modern web technologies} (HTML5, JavaScript, D3.js), \change{and the server on} \remove{and } the Django web framework\refurl{{https://djangoproject.com/}}\remove{ on the server side}. \remove{Extracted entities are turned into RDF triples organised through a \emph{\cendari}-specific ontology.} \remove{The NTE}\change{Faceted} browsing and search functionalities are implemented using ElasticSearch, which unifies the exploration of \change{all} resources provided by the project.
\remove{Communication between \cendari services \change{is} implemented as specific Django extensions.} \change{Django}\remove{is well designed for extensions, but it also} provides a rich ecosystem of extensions that we experimented with to support e.g. ElasticSearch and authentication. \remove{Unfortunately, w}\change{As w}e needed a tight integration to communicate precisely between services, \remove{so} we resorted to implementing our own \change{Django extensions for} faceted-search support, indexing and searching with ElasticSearch, access to semantic web platforms, \change{and} support for very large images through an image server. Although the principles behind all these services are well-known, the implementation is always unexpectedly difficult and time-consuming when it comes to interoperability and large datasets.

One example of scalability relates to the faceted search\remove{. 
Faceted search}\change{, which} allows searching a large set of documents through important type of information or ``facets''. A search query is made of two parts: a textual query as search engine support, and a set of facet names and facet values to restrict the search to these particular values. For example, a search can be related to a date range 1914--1915 (the \emph{date} facet with an interval value), and a person such as ``Wilhelm Röntgen'' (a \emph{person} facet restricted to one name), plus a query term such as ``Columbia''. The result of a faceted search is a set of matching documents, \remove{shown with}\change{showing} snippets of text where \remove{the}\change{searched} terms occur, and a list of all \remove{the}facets \change{and all facet values} \remove{defined for}\change{appearing in} the matching documents\remove{, with all the facet values } (or the 10 most frequent \change{facet values} to avoid saturating the user interface). \change{We defined}\remove{\cendari defines} 10 facet types:\remove{ \emph{application} (the name of the application that created the document), \emph{creator} (person who created the document), \emph{date} (date mentioned), \emph{org} (organisation name mentioned), \emph{person} (person name mentioned), \emph{location} (geographical coordinates mentioned), \emph{place} (place name mentioned), \emph{ref} (well-known reference such as ISBN number of URL mentioned), \emph{language} (language name used), and \emph{project} (name of the \cendari project that holds the document).}
\change{ \emph{application} where document was created, person who is the document's \emph{creator}, \emph{language} used, name of the research \emph{project} holding the document, and mentions of dates/periods (\emph{date}), organisation names (\emph{org}), historical persons (\emph{person}), geo-coordinates (\emph{location}), places (\emph{place}) and document identifiers (\emph{ref}) e.g. ISBN, URL.}

ElasticSearch allows defining a structure for searching (called a mapping), and provides powerful aggregation functions to support scalability. For example, for each query, we receive the list of matching geographical locations and dates; if we show one point per result, the visualisations are over-plotted and the communication between the server and the web client takes minutes to complete.  A typical historical project references easily 10,000 locations and thousands of names. Searching over many projects multiplies the number. Furthermore, \cendari also provides reference datasets such as DBPedia that defines millions of locations and dates. Therefore, our faceted search relies on ElasticSearch aggregation mechanisms, \remove{that allows} returning ranges and aggregates.  Locations are returned as geohash identifiers with a count of matches in each geohash area, which \remove{allows}\change{enables} visualising results quickly at any zoom level. Dates are also returned as intervals \remove{and}\change{with the} number of documents for the specified interval.

The NTE fulfils its role of editor, image viewer, and search interface at scale. It currently serves about 3 millions entities and \change{approximately 800,000} documents with a latency around 1--10~seconds depending on the number of users and complexity of the search queries.

\section{Data Integration and Semantic Services}

\change{The primary objectives of the} \emph{Data Integration Platform (DIP)}\remove{primary objectives} were to integrate relevant archival and historical content from disparate sources into a curated repository, to provide tools for describing and integrating hidden archives and to implement semantic services for enquiring and interlinking of content. \change{The DIP}\remove{It} directly or indirectly supports several stages in the historian's research workflow (from Section \ref{sec:workflow}):
selection of sources (\ref{wf:selection}), planning to visit/visiting archives and libraries (\ref{wf:planning}), knowledge organisation and structuring (\ref{wf:struct}), research refinement and annotation (\ref{wf:refinement}), and searching through relevant material. It contributes to \cendari's enquiry environment by offering new ways to discover meaning and perform historical research\cite{CENDARI2015}. \remove{At the same time, }\change{It ensures} that data from external sources remain\change{s} available in the exact format and version in which they have been used for the research in the first place, thus contributing to the reproducibility of the research. It preserves the original data and their provenance information and sustains the derivatives from the data processing, transformation or data modification operations, ensuring data traceability.

To create a representative and rich pool of resources related to the modern and medieval history, the \remove{\cendari}team identified and contacted more than 250 cultural heritage institutions. We encountered a significant diversity among institutions in the level of digitisation and digital presence. Some institutions provide excellent digital access, while others are still in the process of digitising their original analogue finding aids. It should be noted that neither the digital presence itself, nor the existence of digitised material guarantees that the material is publicly available and accessible outside of the institution’s own internal system. Furthermore, differences exist among institutions’ data provision protocols (when available). 

\change{To address these challenges and to access and harvest the content from different institutions, we had to establish a flexible and robust data acquisition workflow, confronting at the same time legal, social and technical challenges as described in detail in \cite{cendariwhitebook2015}}. \change{Our process is consistent with the FAIR data principles, designed \emph{to make data Findable, Accessible, Interoperable, and Re-usable}~\cite{FAIR}. Harvested data is preserved in the original form, enriched during processing (see~\ref{sec:dataapi}), and further interlinked based on the enriched and extracted information. Data is retrievable by a \cendari identifier, along with its data origin e.g. the provider institution or the original identifiers. The NTE and Pineapple provide search and browse functionality for end users, while the Data API exposes data in a machine-readable fashion.  When data is processed or enriched, DIP makes sure that a full log of applied transformations and final outputs are preserved and FAIR.
An exception to some of the principles concerns private research data since we decided to balance transparency and confidentiality for metadata extracted from researchers' notes. Additionally, the FAIRness of the integrated data at large also depends on the original data sources.}

\subsection{Incremental Approach to Data Integration}

\emph{Data} in \cendari originate from archives, libraries, research institutions, researchers or other types of contributors of original content including data created in \cendari. They have the following characteristics in common: variety of data licenses and usage policies; heterogeneous formats, conventions or standards used to structure data; multilinguality and diverse granularity and content quality. In addition, initial prospects suggested that \cendari would have to accommodate data beyond \change{just the} textual and include audio and video materials, thus building an infrastructure with high tolerance for such heterogeneity. 

This scenery leads to the term \emph {data soup}, defined as \emph{``a hearty mixture of objects, descriptions, sources and XML formats, database exports, PDF files and RDF-formatted data''}~\change{\cite{cendariwhitebook2015}}. From a higher level perspective, the \emph{data soup} comprises \emph{raw data}, \emph{knowledge base data}, \emph{connected data} and \emph{\cendari-produced data} which require different data management approaches~\cite{tastedatasoup2015}. A mapping into a common data model (as applied in most data integration approaches) would not be possible \remove{and}\change{or} preferred, for several reasons: \emph{lack of a priori knowledge} about data---plenty of standards or custom data formats; often standard formats were brought in multitude of flavors, sometimes even with contradictory semantics e.g  “creator” was used both as an author of an archival description or as a person who wrote a letter to another person; \emph{non-existence of a widely accepted domain data model}---\remove{\cendari} WWI and medievalist groups had different requirements and perspectives on data granularity. The development of a single, widely accepted new schema (or extension of an existing one) takes time and does not guarantee flexibility for future modifications of the schema and the tools\remove{ based on that schema}. New data acquisitions may impose new changes in the metadata schema, which, apart from modifying the tools, causes further delays to the data integration scenarios. It also increases the risk of incompleteness, as data would be limited only to the structure supported by a single schema, \remove{some}\change{thus a} resource not fitting the schema would have to be omitted. 

Even if \cendari would have established an all-inclusive new metadata schema, it would still have not guaranteed that it will serve researchers’ needs,  ensure their transnational and interdisciplinary engagement, and provide an enquiry-savvy environment. Such a schema would either be highly generic and comprehensive, thus defeating the purpose of having a schema; or it would be too specific, thus failing to fulfil the needs of both current and future researcher groups.  

Consequently, it was necessary to adapt a lean model to the data integration, \change{avoiding thorough domain modelling until a better understanding about user requirements and scenarios develops.} The resulting system should enable integration of data processing components in a \textit{pay-as-you-go fashion}~\cite{Hedeler2013}, deliver early results and corrective feedback \remove{to}\change{on} the quality and consistency of data, perform refinement and data enrichment in an incremental rather than prescribed way. 

For this purpose we adopted an approach combining two conceptual frameworks: \textit{dataspace}~\cite{Dataspaces} and \textit{blackboard}~\cite{Blackboard}. This allowed us to separate the data integration process from the data interpretation and developments of domain specific application models and tools \cite{tastedatasoup2015,cendariwhitebook2015}.
The \textit{dataspace} promotes coexistence of data from heterogeneous sources and a data unification agnostic to the domain specifics.
\remove{Although this is often seen as a dirty or inconsistent approach,}Such a system contributes to creating and inferring new knowledge, and benefits from the ``dirt in data''~\cite{YoakumStover2010} by preserving the information in its original form, enabling serendipitous discoveries in data.
The \textit{blackboard} architecture is often illustrated with the metaphor of a group of specialists working on a problem~\cite{Blackboard}, sharing their solutions to the ``blackboard'' and reusing them to develop further solutions \remove{to the problem} until the problem is solved. We applied a slightly modified blackboard model, \remove{where }a specialist \change{being} a data processing service (agent) triggered by \change{a data} insertion or update\remove{ of data}; the processing service produces an output which may be used by other services \remove{or may be sent to one of \cendari's persistence}\change{and} components; additionally, it specifies the certainty score of the output, which is\remove{ then} used\remove{ by} to filter out the less optimal results.

\subsection{Data Integration and Processing Components}\label{sec:dataapi}

\begin{figure}
\centering
\includegraphics[width=0.8\textwidth]{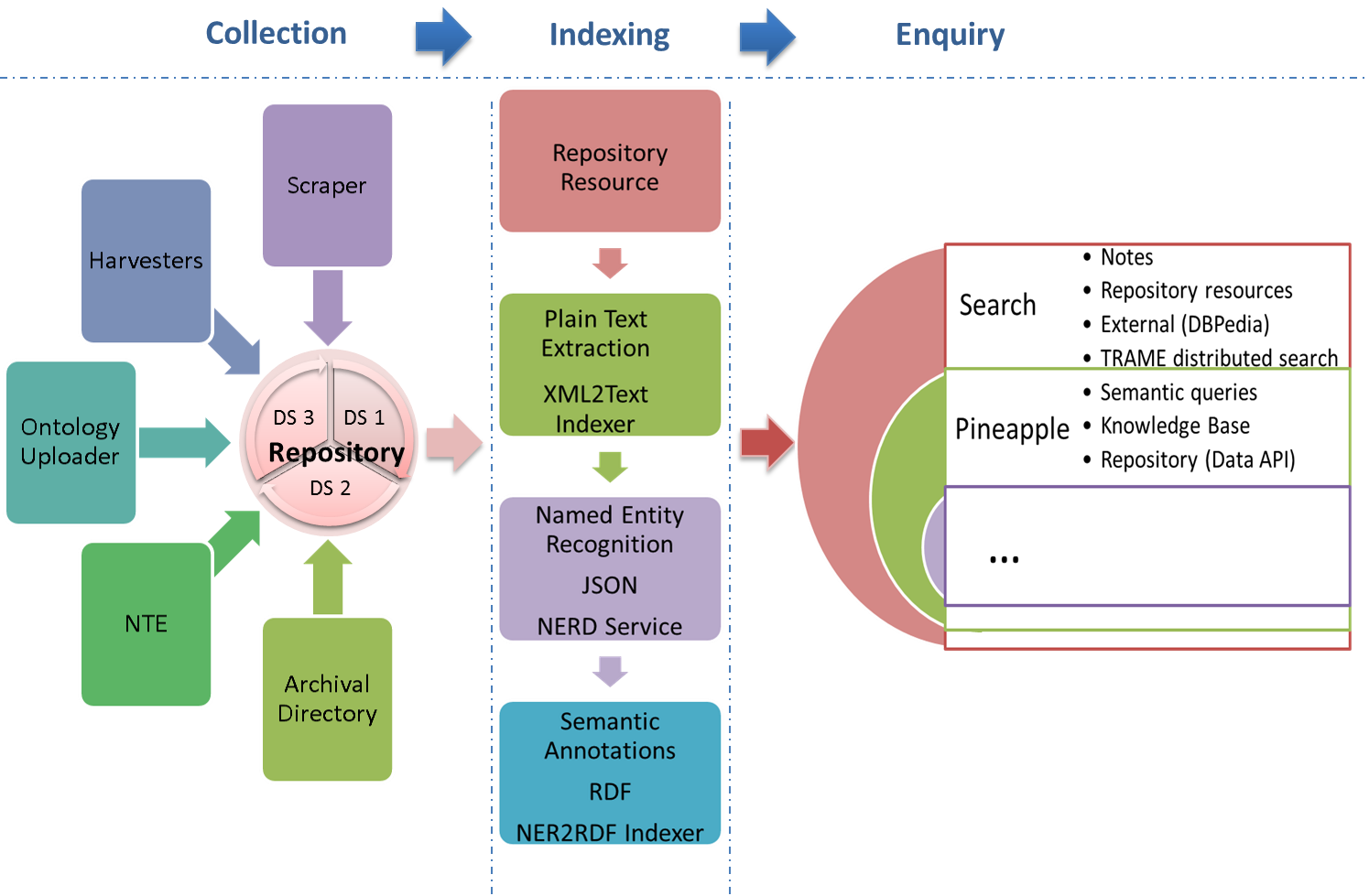}
\caption{Data integration workflows 
         within the \cendari infrastructure \cite{cendariwhitebook2015} }
\label{fig:Cendariintegrationlayers}
\end{figure}

\cendari workflows are grouped around three major processes: collection, indexing and enquiry (Fig. ~\ref{fig:Cendariintegrationlayers})\cite{cendariwhitebook2015}. 
The \remove{data integration platform}\change{DIP} is essential for the data collection and indexing and integrates \remove{several}services to transform and process data for searching and semantic enquiries by the end users \cite{CendariD74}. 

The \change{\textit{Repository}} is a central storage for harvested data and \cendari-created content. It assigns basic provenance information to the data and ensures that data is versioned. In addition\change{,} it keeps the overall authorisation information at a single place. The \remove{\cendari}Repository implements the dataspace model and has no understanding of the data semantics.

The \textit{Data acquisition components} support the collection process by harvesting and ingesting content into the Repository. They vary from dedicated \change{harvesters}\remove{clients (harvesters)} for content providers APIs, to a Scraper service, which extracts structured data directly from the web pages of selected providers, in absence of other data interfaces. Additionally, a special data synchronisation component was developed in order to transfer the Archival Directory data into the Repository.

The \emph{Knowledge Base (KB)} is the central node for \cendari's domain knowledge, based on \change{the} Virtuoso triple store. While Repository data are file-based resources, the KB is where historical knowledge, formalised through ontologies, vocabularies or user annotations, is persisted in a schema-free, but still structured form. The KB is populated from several strands: acquisition of existing structured domain \change{ontologies}\remove{knowledge in the form of domain ontologies},
acquisition of knowledge from collected non-structured/non-RDF resources through semantic data processing, or via the NTE user annotations. The KB \change{supports the dataspace model, implements a RDF named graph for each resource}\remove{is aware of the dataspace model, contains a named graph for each repository resource}, and provides authorisation features 
during the research refinement, annotation and knowledge organisation processes.
The KB is \change{accessed} through the NTE and\remove{the} Pineapple.

The \textit{Data API}\refurl{https://docs.cendari.dariah.eu/developer/litef-conductor/docs/rest.html}  is a\remove{ JSON-based} REST interface which provides machine-readable access to all data in the repository accordingly to the user permissions. It is agnostic towards semantically rich descriptions and is primarily aware of the dataspace, provenance, authorisation and data format. The \remove{\cendari} Data API in addition provides unified access to all derivatives of the original data in the repository, generated through the processing services. 

\textit{Data processing components} enlist a document dispatching service and several internal or external services for resource indexing, acting as a modified blackboard system. These are the Litef Conductor, TikaIndexer, Named Entity Recognition and Disambiguation (NERD) services, Semantic processing services, VirtuosoFeeder and ElasticFeeder services. The following sections provide a short overview of these services.

\subsubsection{Litef Conductor \change{(Litef)}}\label{litef}

\change{Originally named LIve TExt Framework, Litef} implements a dispatching mechanism for invocation of integrated data processing services, either developed by third parties or by \cendari. \change{It separates} the (general or domain specific) data processing from the internal organisation of data, authentication and access permissions and \change{avoids} deep internal dependencies between participating data processing services.  
Litef reacts on the addition or update of a resource in the Repository and passes it to all interested \remove{parties (}data processing services. The results of the processing and their log are stored in the file-system and available via the Data API in read-only mode.

For each data processing service Litef implements a small plug-in, an indexer, which informs Litef about the types of resources it is interested in and the type of result it produces. This is a simple concept, but is expressive enough to define even complex processing flows. The plugin-based indexing architecture ensures that the system can be extended with new processing services, either to support additional formats, or to perform other specific data processing in the future.

\subsubsection{Plain-Text Indexing, Metadata Extraction, Indexing for Search}
Common processing steps for most of the resources are plain-text extraction and metadata extraction. These have been integrated in Litef as \change{a standalone Java library} named TikaExtensions\refurl{https://github.com/CENDARI/tika-extension}, based on the Apache Tika toolkit\refurl{https://tika.apache.org/}\remove{\footnote{developed by \cendari, see \url{https://github.com/CENDARI/tika-extensions}; based on the Apache Tika toolkit}}. The library implements dedicated parsers for most frequent and preferred formats\footnote{EAD/XML, EAG/XML, EDM RDF, METS/XML, MODS/XML, OAI-PMH records, TEI/XML}. 
For other formats, default parsers were used, providing less precise extraction. Litef transforms \change{the} parsed output from the library and generates several resources: a \textit{plain-text} file containing extracted textual selection, a \textit{key-value pairs} file serialising the output from the library, 
and a \textit{search index document} in JSON format, in accordance with the defined search facets (see \ref{sec:nte-technologies}). Where possible, a link to the original location of the resource \remove{was also}\change{is} provided. The ElasticFeeder service then recognises newly generated index document and sends it to the ElasticSearch service, integrated by the NTE, enabling search across all resources, independent from the tool where the resource was originally created.

This approach allows us to separate the metadata extraction logic from Litef internals and to iteratively improve \remove{metadata extraction}\change{it} in a \textit{pay-as-you-go} fashion, as our knowledge about data developed. Furthermore, it allows us to reuse a wide variety of already available Tika-based metadata extraction plugins, to customise them\remove{as needed} or integrate \remove{completely }new parsers. Note that the library can be easily reused outside of the \cendari context.

\subsubsection{NERD Services}

The participatory design workshops showed that historians are mostly interested in identifying places, persons, events, dates and institutions in archival material, annotating them as entities and linking them to the resources where these terms originally appeared. 
To support the entity identification over a large corpus of data, the NERD service was used for automatic entity extraction from pre-processed plain text content. \remove{In the \cendari project t}\change{Two} NERD services were developed \change{in the project}, one for English language~\cite{NERD} and another for multiple languages (Bulgarian, German, Greek, English, Spanish, Finnish, French, Italian, Ripuarisch Platt, Latin, Dutch, Serbian (Cyrillic), Swedish)~\cite{mner}. Both services expose very similar REST APIs and provide JSON-formatted outputs of the recognised entities and the confidence of the result. \change{Although their entity recognition methods vary, they both use Wikipedia for entity disambiguation.} 

\subsubsection{Semantic data extraction and processing}\label{sec:semanticext}

 For each resource in the repository, Litef creates a semantic representation as a named document graph \remove{. E}\change{with e}xtracted metadata added as properties of the resource within that\remove{ document} graph. Depending on the processing results of the NERD services, semantic representations will be created for entity resources of type person, place, event, and organisation, following the \cendari ontology structures. For the most common formats, such as EAD and EAG, more accurate semantic mapping is performed. All outputs of \change{the} semantic processing services are persisted in the \change{KB}.

\subsection{\change{The development of \cendari Knowledge Base (KB)}}\label{sec:semantickb}

\change{A number of ontologies (see conceptualisations in \citep{gruber1993}) were used within the \cendari Knowledge Organisation Framework developments. These vary from metadata schema for describing archival institutions, through controlled vocabularies, gazetteers, to domain ontologies structuring knowledge about both supported research domains.} 

\change{We created extensive ontology development guidelines~\citep{deliverable6.3}, focusing on the reuse of \textit{an existing ontology element set} and \textit{suitable instances} for each domain. In a joint workshop, researchers from both domains identified similar types of concepts and entities to be represented, broadly fitting within the Europeana Data Model (EDM) classes\refurl{http://pro.europeana.eu/page/edm-documentation}: Agent, Place, Timespan, Event, and Concept. Domain differences could be accommodated through EDM extensions, allowing finer level of granularity whilst enabling unification at a courser level of granularity and fostering the data interoperability. 
For example, to coincide with the $100^\textrm{th}$ anniversary of the start of WWI, many research projects published WWI data. However the format and the quality of this data varied considerably:  Excel spreadsheet, data scraped from web pages, and as a badly-formed single large RDF file. We implemented custom solutions to transform relevant existing vocabularies into appropriate EDM extension. Transformed ontologies were aligned to provide better integrated coverage of the domain than was provided by any single ontology on its own. Due to the nature of data, we used a terminological approach to the alignment (for other approaches see \citep{shvaiko2013}), more specifically a character-based similarity measure and the I-SUB technique \citep{Stoilos2005}. The concepts in the ontologies also made use of MACS \citep{ClavelMerin2004MACS} to facilitate multilingual access to the resources for English, German and French.}

\change{Transformed ontologies form a large part of the KB, along with the data automatically generated by the DIP. Data can be browsed and searched through Pineapple, providing a unified view over transformed ontologies, the semantic data extracted from heterogeneous archival resources and, medieval manuscript ontologies, including visual navigation through the text and the structure of the medieval manuscripts. As an example, for an entry about ``Somme'', Pineapple will deliver entries from DBpedia, WWI \change{Linked Open Data\refurl{http://www.ldf.fi/dataset/ww1lod/})}, and Encyclopedia 1914-1918\refurl{https://docs.cendari.dariah.eu/user/pineapple.html} ontologies\refurl{{https://repository.cendari.dariah.eu/organization/cendari-ontologies}}. By navigating to one of them (e.g. the ``Battle of Somme'' event), more information about the event and potentially associated archival resources is displayed. For the latter, more details are available, such as the extracted text from the resource raw data, generated mentions of organisations, places, periods, persons or other semantically related archival resources.} 

\change{Transformed domain ontologies were published in the Repository with the \emph{Ontology Uploader}, which creates additional provenance metadata, based on the OAI-ORE resource maps model\refurl{{http://www.openarchives.org/ore/1.0/datamodel}}, expressing the relationships between the original and transformed data.} 

\change{Element set ontologies developed or adapted within the project are published in raw format on GitHub~\cite{Cendari:dev} and available for visual exploration through the WebVowl tool~\refurl{http://vowl.visualdataweb.org/webvowl.html}. A smaller portion of the KB was created by researchers, through the NTE. They were primarily focused on identifying and tagging entities within notes and uploaded documents, and resolving them against DBPedia (see ~\ref{sec:nte}). This knowledge was captured according to the schema.org general purpose ontology built into the original software used for the NTE. Provisioning of a richer semantic editor or annotator tool to support user friendly ontology developments by researchers, along with relations between entities, notes and any other archival resources, proved to be very challenging. Further developments required strong balance between the flexibility of the tool and the simplicity of the user interface, which was deemed beyond the scope of the \cendari project.}

\section{Discussion}

The goal of the \cendari project was to support historical researchers in their daily work. All the infrastructure described has been designed with that goal in mind. This goal is relatively original in the digital humanities landscape and required a lot of experiments and a complicated infrastructure. Although the goal has been reached technically, we only know at the end of the project what the essential components needed are and the capabilities and digital literacy of historians in general.  We try to summarise here the positive outcomes of the project and some of the pitfalls.

It is very clear at the end of the project that historians benefit immensely from supporting tools such as the one offered by \cendari. Data management at the historian level has become a technical nightmare without proper support. As discussed in Section~3, \cendari benefited from the feedback offered by historians to express their workflows and needs; to our knowledge, they were never explicitly stated before in such a constructive way. Even if our workflow does not support all the research steps performed by historians, it supports a large portion of them.

The infrastructure to support faceted search and semantic enrichment is very complex. Is it worth the effort when large companies such as Google and Microsoft are investing in search engines?  Our answer is positive: historians need more specific support than what search engines currently offer. It may sound like a banality, but here is the right place to state that not all data is digital; rather the opposite is true if cultural heritage is being taken into account. Unique historical data exists only in paper format. The largest part of historical material is neither digitised nor available via metadata. While search engines are targeted at modern-life resources, historians want information on past time-ranges. They also have to deal with places that changed their name or even disappeared. Modern search tools are not meant for these goals and should be supplemented by more focused tools like the ones we designed. Our ingestion and enrichment tools are complex because they need to deal with multiple types of data, and extract useful information out of them \change{in order} to be usable by historians. Offering a faceted search engine is very useful to historians because it helps them contextualise the search results through time, location, etc. However, extracting good quality entities has a cost that explains the complexity of \cendari. In addition to the automatic extraction of entities from existing documents and data, it is now clear that a large portion of archive and library data will never get digitised, as acknowledged by all the institutions that the \cendari project interacted with. Therefore, \cendari has decided to use the data gathered by researchers as historical resource; we believe that this decision is essential to better support historical research in general. We believe that the ``political'' decision to keep the historians' notes private by default but to publish the tagged entities publicly is an effective way to spread and share historical knowledge with little risk of disclosing historians' work.

We realised that there is a tension between the \remove{power}\change{capabilities} technology can provide and the digital training of historians to understand and use these \remove{powerful} capabilities. Our user interface supports some powerful functions simply, but there are limitations. For example, allowing manual or automatic tagging of notes and transcriptions was perceived as very useful and important, but we only support seven types of entities because the user interface would become very complicated if we needed to support more.  Allowing more complex enrichment is possible using the RDFace editor that we provide, but has seldom been used because of the complexity of manually tagging text through ontologies.

We gathered knowledge about how researchers work and some typical workflows. We documented all these and started a co-evolution between our tools and the historians' capabilities and needs, but more iterations became necessary: our agile development processes\change{, as outlined in section~\ref{subsec:iterativeapproach},} allowed us to perform short iterations, \change{allowing code deployments to production in mere hours before evaluation workshops.} 
But \change{conversely this added additional complexity at every step}.
On the other hand, some level of simplification is always needed to reach a new target audience, such as historian researchers.
Early integration of the infrastructural components is essential to ensure timely feedback and reduce friction later, but in a setup of parallel and distributed development efforts with several teams on individual\remove{agile iteration} \change{development iteration} schedules, efficient and clear communication among all participants is a crucial factor to align the \remove{efforts}\change{work} and create a common and collaborative development effort.

\change{By mixing agile methods and long term planning, \cendari built a reproducible infrastructure that has since been taken over by DARIAH in an effort to ensure its sustainability and availability for future use by historians and scholars from other disciplines alike.}

\remove{While agile development can foster these processes, after reaching some level of complexity, long term planning is important too.
Early integration of the infrastructural components is essential to ensure timely feedback and reduce friction later. But in a setup of parallel and distributed development efforts with several teams on individual agile iteration schedules, efficient and clear communication among all participants is a crucial factor to align the efforts and create a common and collaborative development effort.

Our agile development methods and DevOps approach enabled us to reach the development goals in short iterations, allowing code deployments to production in mere hours before evaluation workshops.
At the same time they added another level of complexity to the technical development as well.
Nethertheless, by leveraging the existing DARIAH services and applying agile development methods to the underlying infrastructure building, we built a reproducible system that can be recreated on demand from code.
This tightly integrated environment has since been taken over by DARIAH in an effort to ensure its sustainability and availability for future use by historians and scholars from other disciplines alike.}

\section{Conclusion}


Through the functional requirements identified during the Participatory Design Workshops, a firm basis could be laid in order to support historians in performing their specific workflow. The tools and services provided by the \cendari research infrastructure favor ordering, organisation, search, taking notes, annotations, transcribing, but also the analysis of networks and time-space-relationships as well as sharing of resources. Researchers are thus supported in drawing together resources and facts about\remove{ the} persons, organisations and structures in the timeframe under consideration. These can be pulled together in order to make patterns visible which cannot be easily taken into focus by classical historiographical methods.

An interesting result of the \cendari project was the formulation of requirements by historians which \remove{do not immediately}\change{support not just their specific research workflow, but rather} the research process as a whole\remove{ as well as their working environment}. The \remove{implemented} visualisations and the built-in collaboration functionalities of the \cendari infrastructure -- like sharing of resources, the establishment of collaborative projects or the possibility of collaborative writing of articles -- seem at first glance secondary to the research process, but enhance the analysis of search results and the community of historians in general. This can be seen as the ``pedagogical'' offer of the infrastructure.  While historians are generally trained to work all by themselves, the infrastructure offers a range of possibilities for collaborative information collection and writing. It thus lays the basis for a truly collaborative and transnational historiography.
 
Furthermore, the examples resulting from the collaboration of information engineers, historians and archivists are very promising beyond the achievement of the \cendari project. The development of ontologies and the possibility of their collaborative enlargement by users can be regarded as a potentially fruitful domain for interaction between the disciplines involved. Another example is the enrichment of metadata by researchers in international standard formats and their hand-over to the cultural heritage institutions which established and provided the metadata. Quite obviously an important part of historians’ work in archives consists of a deeper description of archival material than the one provided by archivists and librarians, who aim at a much more formal level of description. Provided the interoperability of the data through the infrastructure, enriched metadata shared by users and cultural heritage institutions can be described as a win-win-situation for all the sides involved. \change{To achieve a cross-domain level of interoperability of data and services, however, syntactic conformance and semantic data \emph{per se} are not sufficient. Enabling tools for researchers to structure their knowledge and map it across different domains calls for joint efforts in the domain modelling, technical implementation across research infrastructures, training and communication with researchers and strong research community participation. Could a medieval archaeologist working with e.g. ARIADNE\refurl{http://www.ariadne-infrastructure.eu/} benefit from \cendari? Our answer is positive, but we are aware that, additional service-level integration or semantic-data level and ontology developments alignment would be needed for flawless user experience. }

\remove{Nevertheless, the}\change{The} infrastructure built by the \cendari project does not support several steps of classical hermeneutic interpretation which is typical for historians. It can be questioned whether there will ever be tools to support humanists in the specific practises and competences which mark this profession --- the observation and interpretation of ambivalence and polysemies, of ambiguities and contradiction, and the differentiated analysis of cultural artifacts. The broad range of tools, services and resources offered by the \cendari \remove{project}\change{infrastructure} underlines the fact that not every need formulated by historians can be satisfied and a mission creep with respect to requirements has to be avoided.

\section{Acknowledgments}
\change{
The research leading to these results has received funding from the European Union Seventh Framework Programme ([FP7/2007-2013] [FP7/2007-2011]) under grant agreement n\textdegree  284432.}

\change{We are thankful to all \cendari data providers who contributed with their content and made it available for research.}

\change{We would like to express our sincere gratitude to all \cendari partners for their great contributions to the development and setup of the infrastructure. The fusion of researchers, archivists, librarians and IT experts had made the \cendari project a unique learning experience for all of us. }

\bibliographystyle{ACM-Reference-Format-Journals}
\bibliography{references}


\begin{thebibliography}{00}


\ifx \showCODEN    \undefined \def \showCODEN     #1{\unskip}     \fi
\ifx \showDOI      \undefined \def \showDOI       #1{{\tt DOI:}\penalty0{#1}\ }
  \fi
\ifx \showISBNx    \undefined \def \showISBNx     #1{\unskip}     \fi
\ifx \showISBNxiii \undefined \def \showISBNxiii  #1{\unskip}     \fi
\ifx \showISSN     \undefined \def \showISSN      #1{\unskip}     \fi
\ifx \showLCCN     \undefined \def \showLCCN      #1{\unskip}     \fi
\ifx \shownote     \undefined \def \shownote      #1{#1}          \fi
\ifx \showarticletitle \undefined \def \showarticletitle #1{#1}   \fi
\ifx \showURL      \undefined \def \showURL       #1{#1}          \fi

\bibitem[\protect\citeauthoryear{{Beaudouin-Lafon} and
  Mackay}{{Beaudouin-Lafon} and Mackay}{2002}]%
        {beaudouin-Lafon:2002}
{Michel {Beaudouin-Lafon}} {and} {Wendy Mackay}. 2002.
\newblock \showarticletitle{Prototyping Development and Tools}.
\newblock In {\em Human Computer Interaction Handbook}, {Julie~A. Jacko} {and}
  {Andrew Sears} (Eds.). Lawrence Erlbaum Associates, Hillsdale, NJ, USA,
  1006--1031.
\newblock
\showURL{%
\url{http://www.isrc.umbc.edu/HCIHandbook/}}


\bibitem[\protect\citeauthoryear{Becker and Cleveland}{Becker and
  Cleveland}{1987}]%
        {brushing}
{Richard~A. Becker} {and} {William~S. Cleveland}. 1987.
\newblock \showarticletitle{Brushing Scatterplots}.
\newblock {\em Technometrics\/} {29}, 2 (1987), 127--142.
\newblock


\bibitem[\protect\citeauthoryear{Blanke and Kristel}{Blanke and
  Kristel}{2013}]%
        {blanke2013integrating}
{Tobias Blanke} {and} {Conny Kristel}. 2013.
\newblock \showarticletitle{Integrating holocaust research}.
\newblock {\em International Journal of Humanities and Arts Computing\/} {7},
  1-2 (2013), 41--57.
\newblock


\bibitem[\protect\citeauthoryear{Boukhelifa, Giannisakis, Dimara, Willett, and
  Fekete}{Boukhelifa et~al\mbox{.}}{2015}]%
        {boukhelifa2015}
{Nadia Boukhelifa}, {Emmanouil Giannisakis}, {Evanthia Dimara}, {Wesley
  Willett}, {and} {Jean-Daniel Fekete}. 2015.
\newblock \showarticletitle{{Supporting Historical Research Through
  User-Centered Visual Analytics}}. In {\em EuroVis Workshop on Visual
  Analytics (EuroVA)}, {E.~Bertini} {and} {J.~C. Roberts} (Eds.). The
  Eurographics Association, Calgiary, Italy.
\newblock


\bibitem[\protect\citeauthoryear{CENArch}{CENArch}{2015}]%
        {ADT}
CENArch 2015.
\newblock CENDARI Archival Directory.
\newblock   (2015).
\newblock
\showURL{%
\url{https://archives.cendari.dariah.eu/}}


\bibitem[\protect\citeauthoryear{CENDARI}{CENDARI}{2015}]%
        {CENDARI2015}
CENDARI 2015.
\newblock Collaborative European Digital Archive Infrastructure.
\newblock   (2015).
\newblock
\showURL{%
\url{www.cendari.eu}}


\bibitem[\protect\citeauthoryear{CENGitHub}{CENGitHub}{2016}]%
        {Cendari:dev}
CENGitHub 2016.
\newblock Cendari Development Repository on GitHub.
\newblock \url{https://github.com/cendari/}.   (2016).
\newblock
\newblock
\shownote{Accessed: 2016-04-26.}


\bibitem[\protect\citeauthoryear{CENOnt}{CENOnt}{2014}]%
        {deliverable6.3}
CENOnt 2014.
\newblock Deliverable 6.3: Guidelines for Ontology Building.
\newblock   (2014).
\newblock
\showURL{%
\url{http://www.cendari.eu/sites/default/files/CENDARI_D6.3GuidelinesforOntologyBuilding.pdf}}


\bibitem[\protect\citeauthoryear{CENTools}{CENTools}{2016}]%
        {CendariD74}
CENTools 2016.
\newblock Deliverable D7.4: Final releases of toolkits.
\newblock   (2016).
\newblock
\showURL{%
\url{http://www.cendari.eu/sites/default/files/CENDARI_D7.4-Finalreleasesoftoolkits.pdf}}
\newblock
\shownote{Accessed: 2016-04-27.}


\bibitem[\protect\citeauthoryear{Clavel-Merrin}{Clavel-Merrin}{2004}]%
        {ClavelMerin2004MACS}
{Genevieve Clavel-Merrin}. 2004.
\newblock \showarticletitle{MACS (Multilingual Access to Subjects): A Virtual
  Authority File Across Languages}.
\newblock {\em Cataloging \& Classification Quarterly\/} {39}, 1-2 (2004),
  323--330.
\newblock


\bibitem[\protect\citeauthoryear{Edmond, Beneš, Bulatović, Knežević,
  Lehmann, Morselli, and Zamoiski}{Edmond et~al\mbox{.}}{2015a}]%
        {cendariwhitebook2015}
{Jennifer Edmond}, {Jakub Beneš}, {Nataša Bulatović}, {Milica Knežević},
  {Jörg Lehmann}, {Francesca Morselli}, {and} {Andrei Zamoiski}. 2015a.
\newblock {\em The CENDARI White Book of Archives}.
\newblock {T}echnical {R}eport. CENDARI.
\newblock
\showURL{%
\url{http://hdl.handle.net/2262/75683}}


\bibitem[\protect\citeauthoryear{Edmond, Bulatovic, and O'Connor}{Edmond
  et~al\mbox{.}}{2015b}]%
        {tastedatasoup2015}
{Jennifer Edmond}, {Natasa Bulatovic}, {and} {Alexander O'Connor}. 2015b.
\newblock \showarticletitle{The Taste of “Data Soup” and the Creation of a
  Pipeline for Transnational Historical Research}.
\newblock {\em Journal of the Japanese Association for Digital Humanities\/}
  {1}, 1 (2015), 107--122.
\newblock


\bibitem[\protect\citeauthoryear{Franklin, Halevy, and Maier}{Franklin
  et~al\mbox{.}}{2005}]%
        {Dataspaces}
{Michael Franklin}, {Alon Halevy}, {and} {David Maier}. 2005.
\newblock \showarticletitle{From Databases to Dataspaces: A New Abstraction for
  Information Management}.
\newblock {\em SIGMOD Rec.\/} {34}, 4 (Dec. 2005), 27--33.
\newblock
\showISSN{0163-5808}


\bibitem[\protect\citeauthoryear{Gruber}{Gruber}{1993}]%
        {gruber1993}
{Thomas~R. Gruber}. 1993.
\newblock \showarticletitle{A Translation Approach to Portable Ontology
  Specifications}.
\newblock {\em Knowl. Acquis.\/} {5}, 2 (June 1993), 199--220.
\newblock
\showISSN{1042-8143}


\bibitem[\protect\citeauthoryear{Hayes-Roth}{Hayes-Roth}{1985}]%
        {Blackboard}
{Barbara Hayes-Roth}. 1985.
\newblock \showarticletitle{A Blackboard Architecture for Control}.
\newblock {\em Artif. Intell.\/} {26}, 3 (Aug. 1985), 251--321.
\newblock
\showISSN{0004-3702}


\bibitem[\protect\citeauthoryear{Hedeler, Fernandes, Belhajjame, Mao, Guo,
  Paton, and Embury}{Hedeler et~al\mbox{.}}{2013}]%
        {Hedeler2013}
{Cornelia Hedeler}, {Alvaro A.~A. Fernandes}, {Khalid Belhajjame}, {Lu Mao},
  {Chenjuan Guo}, {Norman~W. Paton}, {and} {Suzanne~M. Embury}. 2013.
\newblock {\em Advanced Query Processing: Volume 1: Issues and Trends}.
\newblock Springer, Berlin, Heidelberg, Chapter A Functional Model for
  Dataspace Management Systems, 305--341.
\newblock
\showISBNx{978-3-642-28323-9}


\bibitem[\protect\citeauthoryear{Iggers}{Iggers}{2005}]%
        {historiography}
{Georg~G. Iggers}. 2005.
\newblock {\em Historiography in the Twentieth Century: From Scientific
  Objectivity to the Postmodern Challenge}.
\newblock Wesleyan University Press, Middletown, CT, USA. 208 pages.
\newblock
\showISBNx{978-0819567666}


\bibitem[\protect\citeauthoryear{Kim, Behr, and Spafford}{Kim
  et~al\mbox{.}}{2013}]%
        {Kim:PhoenixProject}
{Gene Kim}, {Kevin Behr}, {and} {George Spafford}. 2013.
\newblock {\em The Phoenix Project: A Novel About IT, DevOps, and Helping Your
  Business Win\/} (1st ed.).
\newblock IT Revolution Press.
\newblock
\showISBNx{0988262592, 9780988262591}


\bibitem[\protect\citeauthoryear{Lopez}{Lopez}{2009}]%
        {NERD}
{Patrice Lopez}. 2009.
\newblock \showarticletitle{GROBID: Combining automatic bibliographic data
  recognition and term extraction for scholarship publications}.
\newblock In {\em Research and Advanced Technology for Digital Libraries}.
  Springer, Berlin, Germany, 473--474.
\newblock


\bibitem[\protect\citeauthoryear{Mattern, Jeng, He, Lyon, and Brenner}{Mattern
  et~al\mbox{.}}{2015}]%
        {mattern:2015}
{Eleanor Mattern}, {Wei Jeng}, {Daqing He}, {Liz Lyon}, {and} {Aaron Brenner}.
  2015.
\newblock \showarticletitle{Using participatory design and visual narrative
  inquiry to investigate researchers? data challenges and recommendations for
  library research data services}.
\newblock {\em Program: electronic library and information systems\/}  {49}
  (2015), 408--423.
\newblock
Issue 4.


\bibitem[\protect\citeauthoryear{Meyer}{Meyer}{2016}]%
        {mner}
{Alexander Meyer}. 2016.
\newblock Multilingual Named Entity Recognition and Resolution.
\newblock   (2016).
\newblock
\showURL{%
\url{http://136.243.145.239/nerd/}}


\bibitem[\protect\citeauthoryear{Muller}{Muller}{2003}]%
        {muller:2002}
{Michael~J. Muller}. 2003.
\newblock \showarticletitle{Participatory Design: The Third Space in HCI}.
\newblock In {\em The Human-computer Interaction Handbook}, {Julie~A. Jacko}
  {and} {Andrew Sears} (Eds.). Lawrence Erlbaum Associates, Hillsdale, NJ, USA,
  1051--1068.
\newblock
\showISBNx{0-8058-3838-4}
\showURL{%
\url{http://dl.acm.org/citation.cfm?id=772072.772138}}


\bibitem[\protect\citeauthoryear{Shvaiko and Euzenat}{Shvaiko and
  Euzenat}{2013}]%
        {shvaiko2013}
{P. Shvaiko} {and} {J. Euzenat}. 2013.
\newblock \showarticletitle{Ontology Matching: State of the Art and Future
  Challenges}.
\newblock {\em IEEE Transactions on Knowledge and Data Engineering\/} {25}, 1
  (Jan 2013), 158--176.
\newblock
\showISSN{1041-4347}


\bibitem[\protect\citeauthoryear{Stoilos, Stamou, and Kollias}{Stoilos
  et~al\mbox{.}}{2005}]%
        {Stoilos2005}
{G Stoilos}, {G Stamou}, {and} {S Kollias}. {2005}.
\newblock \showarticletitle{{A string metric for ontology alignment}}. In {\em
  {Semantic Web - ISWC 2005, Proceedings}} {\em ({LEcture Notes in Computer
  Science})}, {{Gil, Y and Motta, E and Benjamins, VR and Musen, MA}} (Ed.),
  Vol. {3729}. {Springer-Verlag}, {Berlin, Germany}, {624--637}.
\newblock
\showISBNx{{3-540-29754-5}}
\showISSN{{0302-9743}}
\newblock
\shownote{{4th International Semantic Web Conference (ISWC 2005), Galway,
  IRELAND, NOV 06-10, 2005}.}


\bibitem[\protect\citeauthoryear{Trame}{Trame}{2016}]%
        {TRAME}
Trame 2016.
\newblock TRAME: Text and manuscript transmission of the Middle Ages in Europe.
\newblock \url{http://trame.fefonlus.it/}.   (2016).
\newblock
\newblock
\shownote{Accessed: 2016-04-26.}


\bibitem[\protect\citeauthoryear{Visconti}{Visconti}{2016}]%
        {visconti:2016}
{Amanda Visconti}. 2016.
\newblock Infinite Ulysses.
\newblock   (2016).
\newblock
\showURL{%
\url{http://www.infiniteulysses.com/}}


\bibitem[\protect\citeauthoryear{Warwick}{Warwick}{2012}]%
        {warwick:2012}
{Claire Warwick}. 2012.
\newblock \showarticletitle{Studying users in digital humanities}.
\newblock In {\em Digital Humanities in Practice}, {Claire Warwick}, {Melissa~M
  Terras}, {and} {Julianne Nyhan} (Eds.). Facet Publishing in association with
  UCL Centre for Digital Humanities, London, Chapter~1.
\newblock


\bibitem[\protect\citeauthoryear{Wessels, Borrill, Sorensen, McLaughlin, and
  Pidd}{Wessels et~al\mbox{.}}{2015}]%
        {wessels:2015}
{Bridgette Wessels}, {Keira Borrill}, {Louise Sorensen}, {Jamie McLaughlin},
  {and} {Michael Pidd}. 2015.
\newblock Understanding Design for the Digital Humanities. Studies in the
  Digital Humanities. Sheffield: HRI Online Publications.
\newblock   (2015).
\newblock
\showURL{%
\url{http://www.hrionline.ac.uk/openbook/chapter/understanding-design-for-the-digital-humanities}}


\bibitem[\protect\citeauthoryear{Wilkinson, Dumontier, Aalbersberg, and {\sc
  others}}{Wilkinson et~al\mbox{.}}{2016}]%
        {FAIR}
{Mark~D. Wilkinson}, {Michel Dumontier}, {IJsbrand~Jan Aalbersberg}, {and}
  {{\sc others}}. 2016.
\newblock \showarticletitle{The FAIR Guiding Principles for scientific data
  management and stewardship}.
\newblock {\em Scientific Data\/} {3}, 160018 (2016).
\newblock
\showDOI{%
\url{http://dx.doi.org/10.1038/sdata.2016.18}}


\bibitem[\protect\citeauthoryear{Yoakum-Stover}{Yoakum-Stover}{2010}]%
        {YoakumStover2010}
{Suzanne Yoakum-Stover}. 2010.
\newblock Keynote address "Data and Dirt".
\newblock   (12 2010).
\newblock
\showURL{%
\url{http://www.information-management.com/resource-center/?id=10019338}}
\newblock
\shownote{Talk given at the Top Information Manager Symposium, Dec 21 2010
  [Accessed: 2016 04 27].}


\end{thebibliography}


\end{document}